\documentclass[preprintnumbers,amsmath,amssymb,aps,prl,floatfix,lengthcheck,superscriptaddress]{revtex4-2}

\usepackage{bm}
\usepackage{graphicx}
\usepackage{color}
\usepackage{amsmath}
\usepackage{hyperref}
\usepackage{epstopdf}
\usepackage{upgreek}
\usepackage{mathptmx, textcomp}
\usepackage[latin1]{inputenc}
\usepackage[T1]{fontenc}
\usepackage{array, multirow}
\usepackage[table]{xcolor}
\usepackage{booktabs}
\usepackage{longtable}
\usepackage[normalem]{ulem}

\newcommand{\Rbplus}{Rb$^+$ }
\newcommand{\Rbmolplus}{Rb$_2^+$ }

\newcommand{\commentOut}[1]{}

\begin{document}

\title{Transport of a single cold ion immersed in a Bose-Einstein condensate}
\author{T. Dieterle}
\affiliation{5. Physikalisches Institut and Center for Integrated Quantum Science and Technology, Universit\"{a}t Stuttgart, Pfaffenwaldring 57, 70569 Stuttgart, Germany}
\author{M. Berngruber}
\affiliation{5. Physikalisches Institut and Center for Integrated Quantum Science and Technology, Universit\"{a}t Stuttgart, Pfaffenwaldring 57, 70569 Stuttgart, Germany}
\author{C. H\"olzl}
\affiliation{5. Physikalisches Institut and Center for Integrated Quantum Science and Technology, Universit\"{a}t Stuttgart, Pfaffenwaldring 57, 70569 Stuttgart, Germany}
\author{R. L\"{o}w}
\affiliation{5. Physikalisches Institut and Center for Integrated Quantum Science and Technology, Universit\"{a}t Stuttgart, Pfaffenwaldring 57, 70569 Stuttgart, Germany}
\author{K. Jachymski}
\affiliation{Faculty of Physics, University of Warsaw, Pasteura 5, 02-093 Warsaw, Poland}
\author{T. Pfau}
\affiliation{5. Physikalisches Institut and Center for Integrated Quantum Science and Technology, Universit\"{a}t Stuttgart, Pfaffenwaldring 57, 70569 Stuttgart, Germany}
\author{F. Meinert}
\affiliation{5. Physikalisches Institut and Center for Integrated Quantum Science and Technology, Universit\"{a}t Stuttgart, Pfaffenwaldring 57, 70569 Stuttgart, Germany}

\date{\today}

\begin{abstract}
We investigate transport dynamics of a single low-energy ionic impurity in a Bose-Einstein condensate. The impurity is implanted into the condensate starting from a  single Rydberg excitation, which is ionized by a sequence of fast electric field pulses aiming to minimize the ion's initial kinetic energy. Using a small electric bias field, we study the subsequent collisional dynamics of the impurity subject to an external force. The fast ion-atom collision rate, stemming from the dense degenerate host gas and the large ion-atom scattering cross section, allow us to study a regime of frequent collisions of the impurity within only tens of microseconds. Comparison of our measurements with stochastic trajectory simulations based on sequential Langevin collisions indicate diffusive transport properties of the impurity and allows us to measure its mobility. Our results open a novel path to study dynamics of charged quantum impurities in ultracold matter.

\end{abstract}

\maketitle

Unraveling the microscopic details of transport processes forms a prime challenge to understand macroscopic phenomena in complex many-body systems. Paradigms thereof comprise Drude-type electronic motion, Cooper-pairing in superconductors, or hopping dynamics in lattice or spin systems. The advent of ultracold atomic gases has opened exciting new routes to perform microscopy of transport phenomena in quantum systems \cite{Chien2015,Krinner2015}, even at the level of individual particles by deliberately implanting single impurity atoms in the host gas and monitoring their dynamics in real time. Recent advances in controlling neutral impurities interacting with bulk Bose and Fermi gases, for example, offer means to investigate polaron formation \cite{Schirotzek2009,Kohstall2012,Koschorreck2012,Jorgensen2016}, to trace individual particle collisions \cite{Hohmann2017}, or to study exotic impurity dynamics in reduced dimensions \cite{Palzer2009,Meinert2017}. Neutral impurities interact with the quantum gas via short-range contact interaction. In contrast, hybrid platforms combining neutral atoms with co-trapped ions allow for realizing strongly coupled charged impurities, which interact with the host gas via a comparatively long-range polarization potential \cite{Tomza2019}.

%%%%%%%%%%%%%%%%%%%%%%%%%  Figure 1 - Concept  %%%%%%%%%%%%%%%%%%%%%%%%%
\begin{figure}[!ht]
\centering
	\includegraphics[width=\columnwidth]{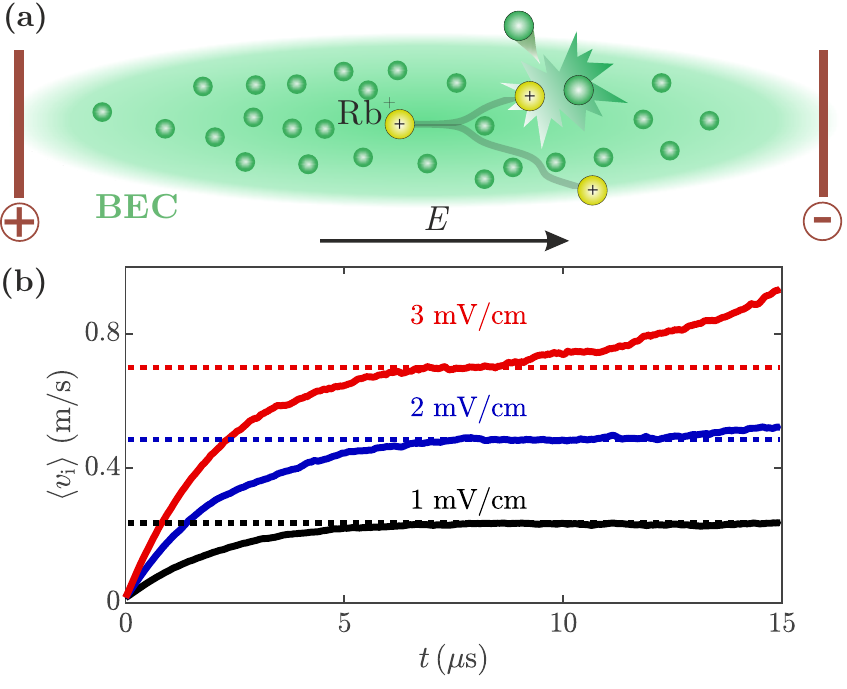}
	\caption{(a) Schematic view of a single \Rbplus ion (yellow sphere) transported through a Bose-Einstein condensate via an applied electric field $E$. Two exemplary ion trajectories indicate frequent binary Langevin scattering with host gas atoms (green spheres) and a competing three-body collision event. (b) Numerical trajectory simulations (see text) of the mean velocity $\langle v_{\rm{i}} \rangle$ as a function of evolution time $t$ for an ion created at the BEC center predict a field-dependent drift velocity (dashed horizontal lines) after an initial transient, indicative of diffusive ion transport.}
	\label{Fig1}
\end{figure}

The strong ion-atom interaction has been exploited not only for sympathetic collisional cooling of the ionic impurity by its host gas \cite{Zipkes2010,Meir2016,Feldker2020,Dutta2018,Schmidt2020}, but also for extensive studies of cold two- and three-body charge-neutral chemistry induced by the long-range tail of the polarization potential \cite{Grier2009,Haerter2012,Ratschbacher2012,Sikorsky2018}. Further, the rich interplay between elastic collisions and chemical processes has been subject to theoretical studies of ionic impurities at ultralow temperatures, predicting coherent formation of exotic mesoscopic molecular ions with large effective masses \cite{Cote2002,Schurer2017},  novel transport processes driven by resonant charge exchange \cite{Cote2000}, or polaronic effects for ion-doped degenerate gases \cite{Astrakharchik2020}. Indeed, mobility studies of ions in liquid Helium have been exploited early on to probe central properties of superfluidity \cite{Reif1960,Rayfield1963}. However, the experimental realization of directed impurity transport in a cold ion-atom hybrid system so far remained elusive and is impeded by the presence of strong confining potentials and typical intrinsic trap-induced micromotion.

In this work, we study transport of a single cold ion through a Bose-Einstein condensate (BEC) on a completely different experimental footing. We control a free and low kinetic energy ionic impurity in a nominally field-free environment. This ion is generated from a single precursor Rydberg atom by a tailored pulsed field-ionization sequence, and strong Ryd\-berg blockade grants the creation of a single charged impurity only \cite{Balewski2013,Kleinbach2018,Engel2018}. Subsequently, we steer the ion through the BEC by applying exquisitely controlled bias electric fields [Fig.~\ref{Fig1}(a)]. This allows us to observe indications of diffusive ionic transport as a result of the large cross section for ion-atom Langevin collisions. 

The interaction between an ion and neutral atoms at long range is well captured by an induced charge-dipole polarization potential of the familiar form $V(R) = -C_4/(2 R^4)$ \cite{Tomza2019}, where $R$ denotes the internuclear separation and $C_4$ the atom's polarizability ($C_4=318.8$ a.u. for Rb \cite{Holmgren2010}). For the typical collision energies considered in this work, it is this long-range tail which determines the Langevin scattering rate $\gamma_{\rm{L}}=2 \pi n_{\rm{at}} \sqrt{C_4/\mu}$ of binary ion-atom scattering events leading to large deflection angles \cite{Cote2000b}. Here $\mu$ is the ion-atom reduced mass and $n_{\rm{at}}$ denotes the atom number density of the host gas. The rate for such Langevin-type collisions can reach up to hundreds of kHz at densities typical for gaseous BECs ($n_{\rm{at}} \sim 10^{14} \rm{cm}^{-3}$), and may thus lead to frequent collisions on timescales of only a few tens of microseconds. Such fast scattering rates form the basis for our experiments and specifically allow us to observe collision-dominated transport dynamics of an untrapped ion controlled by small electric bias fields.

\begin{figure}[!ht]
\centering
	\includegraphics[width=\columnwidth]{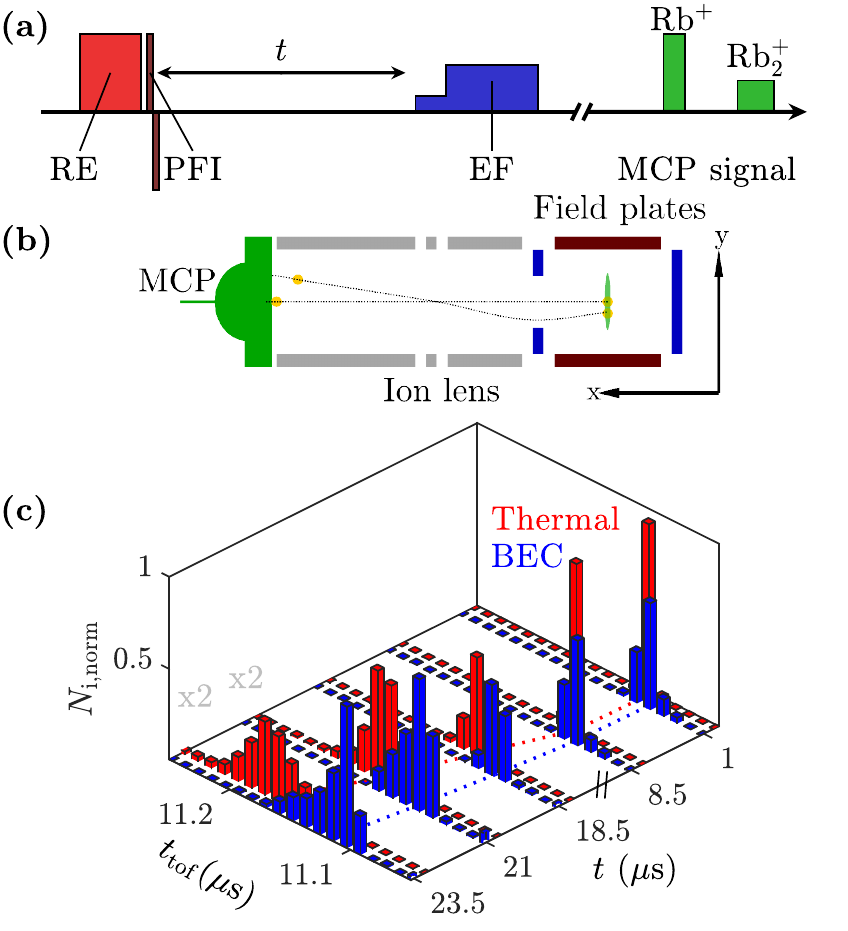}
	\caption{(a) Measurement sequence consisting of the excitation of a single Rydberg atom (RE) and production of a low-energy ion via pulsed field ionization (PFI). After a variable evolution time $t$ in the electric field $E$, a two-step electric field pulse (EF) guides the ion to the MCP for temporarily separated detection of Rb$^+$ and Rb$_2^+$. (b) Detection principal of ionic transport. The two-step extraction pulse (EF in (a)) in combination with an ion lens in einzel lens configuration results in a time of flight to the MCP which depends on the position of the ion in the BEC after time $t$. The color of the electrodes encodes, which field plates are used for applying the PFI and EF pulses shown in (a). (c) Distribution of the ion's time of flight $t_{\rm{tof}}$ as a function of evolution time $t$ for $E=4.3 \, \rm{mV/cm}$ measured in a dense BEC (blue) and a low-density regime for reference (red). The low-density data is slightly offset in $t$ for better visibility.}
	\label{Fig2}
\end{figure}

Our experiments start with a magnetically trapped BEC of typically $8 \times 10^5$ $^{87}$Rb atoms in the  $|5 S_{1/2}, F=2, m_F=2\rangle$ hyperfine state. Trap frequencies of $(\omega_x,\omega_y,\omega_z)=2 \pi \times (194,16,194) \, \rm{Hz}$ yield a condensate with Thomas-Fermi radii of $(4.4,53,4.4) \, \mu\rm{m}$ and a peak density of $n_{\rm{at}} = 4.1 \times 10^{14} \, \rm{cm}^{-3}$. The BEC is surrounded by three pairs of electrodes [Fig.~\ref{Fig2}(b)] allowing to control electric fields at the gas position to a level $\lesssim 300\, \mu\rm{V/cm}$  \cite{Engel2018}. Next, we implant a single low-energy \Rbplus ion into the central region of the condensate by promoting one atom from the atomic ensemble to a highly excited $160S_{1/2}$ Rydberg level with a tightly focused excitation laser ($1.8 \, \mu \rm{m}$ waist) \cite{SM}. Subsequently, the single Rydberg atom is ionized by a short (10~ns long) electric field pulse of 2.85~V/cm followed by a second tailored field pulse of similar shape but with opposite polarity. This sequence allows us to rapidly decelerate the produced ion and results in small initial velocities of $\lesssim 0.1 \, \rm{m/s}$ ($E_{\rm{kin}}\approx k_{\rm{B}} \times 50 \,\mu$K) \cite{SM}.

%%%%%%%%%%%%%%%%%%%%%%%%%  Figure 2 - Experiment  %%%%%%%%%%%%%%%%%%%%%%%%%

The produced low-energy ionic impurity is now accelerated via a deliberately applied electric field $E$ of several mV/cm pointing along the long $y$-axis of the BEC. After a variable evolution time $t$, we interrupt the transport dynamics with a two-step electric field pulse rapidly guiding the ion towards a microchannel plate (MCP) for detection [Fig.~\ref{Fig2}(a) and (b)] \cite{SM}. Finally, the entire procedure is repeated 50 times with the same ensemble of atoms to gain statistics. Along its path to the MCP, the \Rbplus ion passes through an ion lens in einzel lens configuration, resulting in a time of flight $t_{\rm{tof}}$ to the detector which depends on the ion position prior to the extraction sequence [Fig.~\ref{Fig2}(b)]. In some cases, we detect a formed molecular ion \Rbmolplus, which due to its larger mass arrives at much later time at the MCP.

Let us now focus on the distribution of the \Rbplus arrival times $t_{\rm{tof}}$ at the MCP. An exemplary data set for a field $E=4.3 \, \rm{mV/cm}$ and different values of $t$ is shown in Fig.~\ref{Fig2}(c) (blue bars). We compare this measurement to a reference data set taken at about $40$ times smaller densities (red bars), for which the rate of ion-atom collisions ($\gamma_{\rm{L}} \approx 24 \, \rm{kHz}$) is sufficiently small so that the ion motion is fully ballistic. Evidently, one finds that the ion arrives at systematically shorter times at the MCP when it is prepared in and accelerated through the dense BEC. This difference is a result of frequent ion-atom collisions, which we quantify in the following in more detail.

%%%%%%%%%%%%%%%%%%%%%%%%%  Figure 3 - Results transport  %%%%%%%%%%%%%%%%%%%%%%%%%
\begin{figure}[!ht]
	\includegraphics[width=\columnwidth]{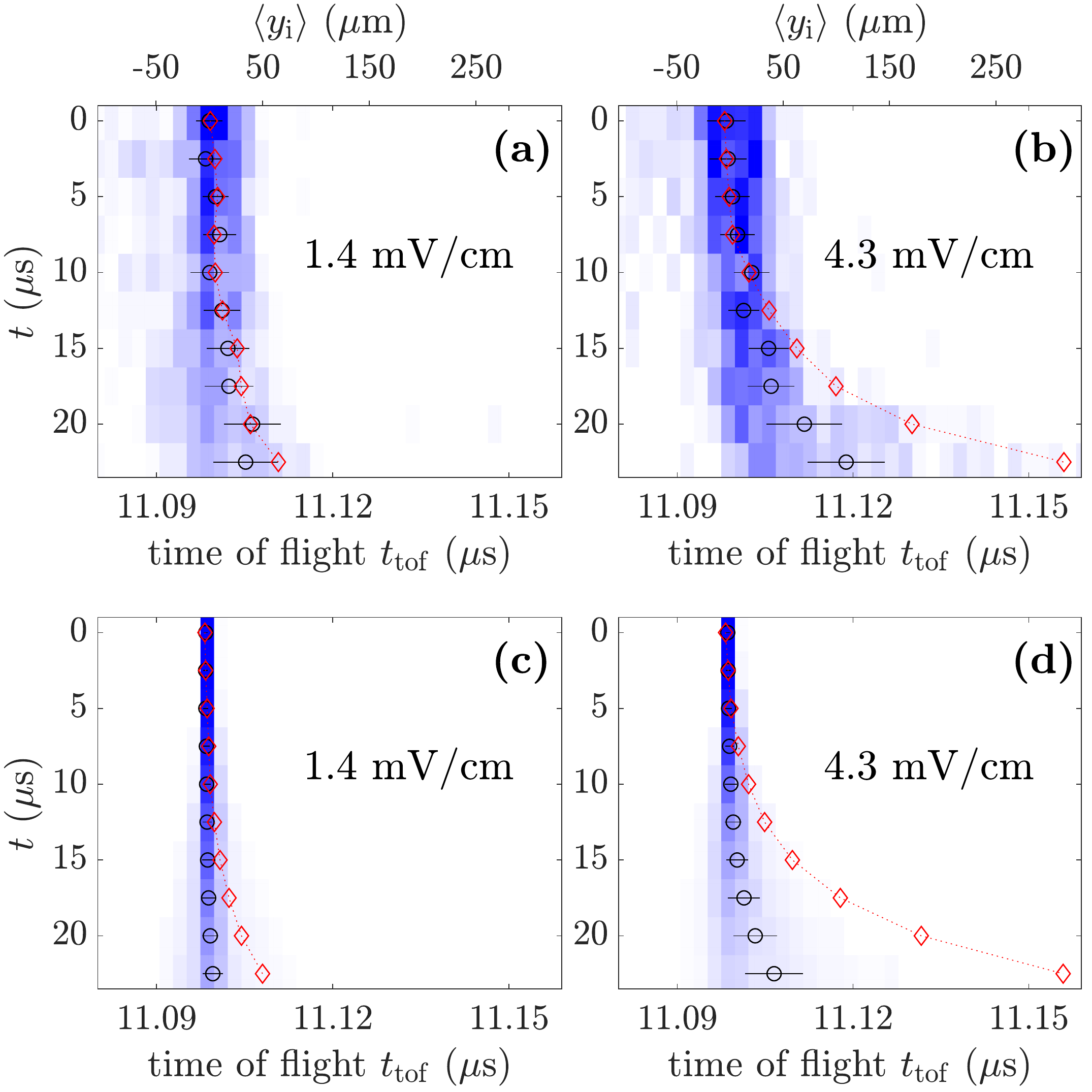}
	\caption{Ion transport through the BEC for an electric field $E=\,1.4\, \rm{mV/cm}$ (a,c) and $E = 4.3\, \rm{mV/cm}$ (b,d). Shown are the measured (top row) and numerically simulated (bottom row) distributions of the ion's time of flight to the MCP as a function of transport time $t$. The color scale encodes the normalized ion signal as shown in Fig.~\ref{Fig2}(c). Black circles depict means of the measured and simulated time of flight distributions $\langle t_{\rm{tof}} \rangle$ and the horizontal bars indicate the corresponding widths at half maximum. For comparison, the red diamonds are measurements and simulations, respectively, of $\langle t_{\rm{tof}} \rangle$ for a ballistic ion dynamics at low density. The upper abscissa in (a) and (b) indicates the mapped mean ion position $\langle y_{\rm{i}} \rangle$ (see text).}
	\label{Fig3}
\end{figure}

To this end, we study the transport dynamics of the ion through the BEC as a function of the applied electric field $E$. Representative measurements showing the distribution of the ion's time of flight $t_{\rm{tof}}$ to the MCP as a function of the transport time $t$ are presented in Fig.~\ref{Fig3}(a) and (b) for two values of $E$. As before, the presence of ion-atom scattering causes systematic shifts of $t_{\rm{tof}}$ towards earlier times compared to reference measurements taken in a dilute collisionless regime of ballistic ion dynamics (red diamonds). For a quantitative ana\-lysis of this observation, we characterize our detection scheme through a detailed analysis of the ion's path to the MCP using a charged-particle trajectory solver \cite{Simion}. For our transport measurements in the BEC, this allows us to map the ion's mean time of flight $\langle t_{\rm{tof}} \rangle$, depicted by the black circles, to its mean axial spatial coordinate $\langle y_{\rm{i}} \rangle$ in the BEC. Note that our analysis accounts for the specific electric-field and ion lens configuration of our setup and is carefully calibrated using the ballistic low-density reference data \cite{SM}. The obtained mean ion position $\langle y_{\rm{i}} \rangle$ is indicated by the upper abscissa for the measurements in Fig.~\ref{Fig3}.
 
Next, we compare our measurements to numerical simulations of the ionic transport dynamics through the condensate. The ion motion is modeled as a series of successive ion-atom collisions with ballistic motion in between \cite{Zipkes2011,Chen2014}. The time for a collision event to happen is Monte-Carlo sampled according to the ion-atom Langevin scattering rate $\gamma_{\rm{L}}$. A second stochastically sampled parameter is the characteristic isotropic scattering angle (in the relative coordinate frame of the colliding ion-atom pair) of spiraling-type Langevin collisions. This stochastic trajectory approach allows us to account for the varying atomic density which the ion probes on its path through the condensate, as well as for the spatial distribution of initial ion positions characterized by the Gaussian profile of our Ryd\-berg excitation laser. Note that we assume only elastic collisions and also neglect the much lower temperature of the atoms in the BEC compared to the average ion kinetic energy. We have also verified that glancing-type forward scattering collisions, which are not accounted for by the Langevin rate, do not significantly affect the transport dynamics \cite{SM}. Numerical results of this model, after a subsequent propagation through our ion-detection setup as described above, deliver the full time-dependent distribution of $t_{\rm{tof}}$. A comparison of the simulated data for the two explored electric field values [Fig.~\ref{Fig3}(c) and (d)] with the experiment yields good agreement. The excellent accordance of the simulated data with the experiment at low densities demonstrates our precise control of the applied electric fields in the range of only a few mV/cm.

%%%%%%%%%%%%%%%%%%%%%%%%%  Figure 4 - time evolution of the mean ion position  %%%%%%%%%%%%%%%%%%%%%%%%%
\begin{figure}[!ht]
\centering
	\includegraphics[width=\columnwidth]{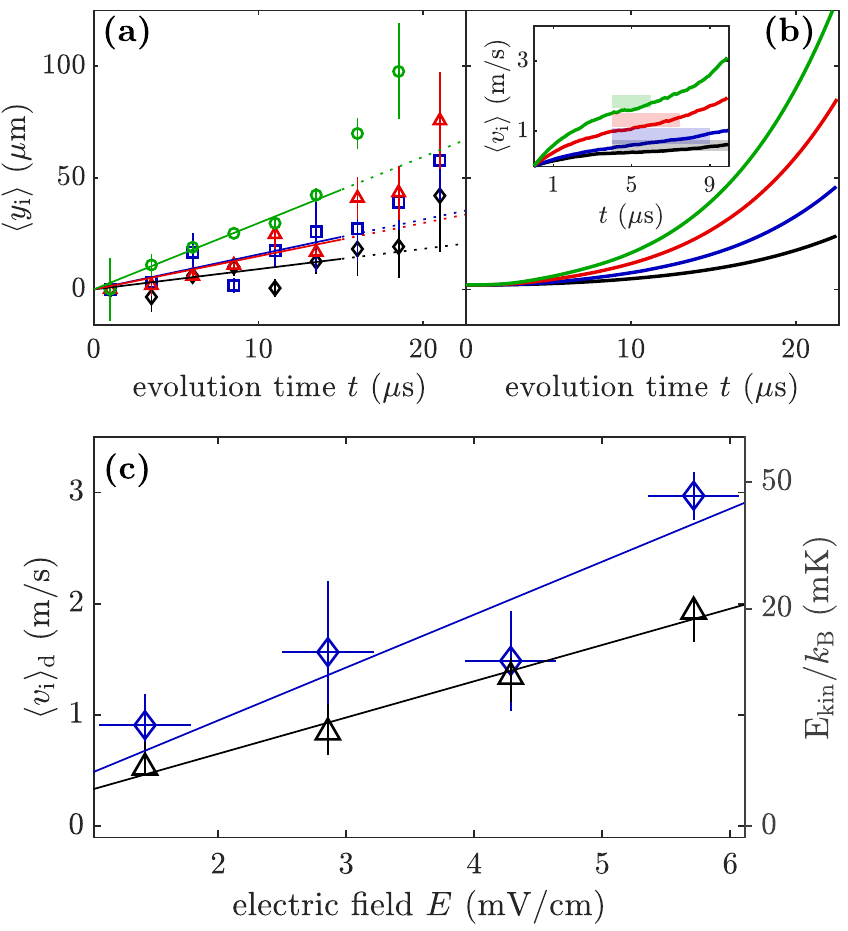}
		\caption{(a) Mean position $\langle y_{\rm{i}} \rangle$ of the ion as a function of evolution time $t$ for $E=1.4$ (black diamonds), $2.8$ (blue squares), $4.3$ (red triangles), $5.7$ (green circles) mV/cm. Error bars originate from the standard deviation of the fitted center of the ion time of flight distribution to its center of mass. Solid lines show linear fits to the data for $t<15$ $\mu$s to analyze the ion's diffusive motion and to extract the associated drift velocity $\langle v_{\rm{i}} \rangle_{\rm{d}}$. (b) Numerically simulated mean ion position $\langle y_{\rm{i}} \rangle$ as a function of evolution time $t$ for the same values of $E$ as in (a), depicted in the same color. The inset shows the corresponding mean particle velocity $\langle v_{\rm{i}} \rangle$. The shaded regions indicate the characteristic plateau of nearly constant transient drift velocity. (c) Measured (blue diamonds) and simulated (black triangles) drift velocity $\langle v_{\rm{i}} \rangle_{\rm{d}}$  as a function of the applied electric field $E$.Vertical and horizontal error bars result from the uncertainty of the linear fits in (a) and indicate the level of stray field control ($\pm$300 $\mu$V/cm) in our setup, respectively. Error bars for the numerical data reflect the shaded regions around $\langle v_{\rm{i}} \rangle_{\rm{d}}$ in the inset of (b). Solid lines are linear fits to the data to extract the ion mobility (see text). The right ordinate shows the kinetic energy of the ion corresponding to $\langle v_{\rm{i}} \rangle_{\rm{d}}$.}
	\label{Fig4}
\end{figure}

Our analysis shows that the reported experiment explores a regime of frequent ion-atom collisions, as a result of which the ionic impurity quickly reaches a state of diffusive transport. This diffusive transport is characterized by a constant drift motion with an electric-field-dependent mean velocity $\langle v_{\rm{i}} \rangle_{\rm{d}}$ [\textit{cf.} Fig.~\ref{Fig1}(b)]. To extract a drift velocity from our measurements, we evaluate the time evolution of the mean ion position $\langle y_{\rm{i}} \rangle$ from data sets as shown in Fig.~\ref{Fig3}. Results for four different values of $E$ are shown in Fig.~\ref{Fig4}(a) together with linear fits to the data, the slopes of which yield values for $\langle v_{\rm{i}} \rangle_{\rm{d}}$. Note that for the case of our finite size condensate, the diffusive motion is a transient state which turns over into ballistic dynamics once the ion approaches the edge of the condensate. For this reason and guided by our numerical calculations, the linear fits for extracting $\langle v_{\rm{i}} \rangle_{\rm{d}}$ exclude data points for $t > 15 \, \mu\rm{s}$. The transient character of the diffusive transport is also apparent in the corresponding simulated time evolution of $\langle v_{\rm{i}} \rangle$ and $\langle y_{\rm{i}} \rangle$ shown in Fig.~\ref{Fig4}(b), which, as mentioned above, takes into account the BEC shape as well as the distribution of the initial ion position. Note that the obtained time evolution of the mean ion position largely deviates from the pure ballistic motion obtained at lower density.

The electric field dependence of $\langle v_{\rm{i}} \rangle_{\rm{d}}$ obtained from the linear fits in Fig.~\ref{Fig4}(a) is shown in Fig.~\ref{Fig4}(c) (blue diamonds). The data follow a linear functional dependence which is a hallmark signature of diffusive transport. From this linear dependence we deduce a mobility for the ionic impurity in our BEC $\mu_{\rm{ion}}^{\rm{exp}} = \partial \langle v_{\rm{i}} \rangle_{\rm{d}}/ \partial E = (47 \pm 16) \times 10^3$ cm$^2$/(Vs).
Finally, the measurements are compared to the prediction from the stochastic trajectory simulation (black triangles), which yields  $\mu_{\rm{ion}}^{\rm{sim}} = (33 \pm 3) \times 10^3$ cm$^2$/(Vs). The numerical results in Fig.~\ref{Fig4}(c) are obtained from linear fits to the simulated data for $\langle y_{\rm{i}} \rangle$ for the same range of $t$ as for the experimental data points, and coincide within $\lesssim 7$\% with drift velocities obtained from the transient plateau in $\langle v_{\rm{i}} \rangle_{\rm{d}}$ [\textit{cf.} inset to Fig.~\ref{Fig4}(b)]. The slightly higher mobility in the experiment compared to the simulations may be attributed to competing inelastic collisions.

Indeed, we observe that in a fraction of the realizations a \Rbmolplus molecule arrives at the MCP instead of a \Rbplus ion. These molecules dynamically form via three-body recombination and can be easily distinguished by their longer time of flight to the detector [\textit{c.f.} Fig.~\ref{Fig2}(a)]. This allows us to exclude events resulting in molecular ion formation from the transport analysis. However, since three-body recombination appears most likely in the densest central part of the excitation volume, our numerical analysis likely overestimates the number of ions stemming from the very center of the BEC and consequently their contribution to the ensemble averaged transport signal. Evidently, such an effect systematically reduces the effective atomic density and qualitatively leads to a larger ion mobility. More quantitatively, we find that for vanishing $E$ and after $t \approx 20 \mu\rm{s}$ about half of the generated ions have been subject to three-body recombination. This number systematically drops for increasing values of $E$ \cite{SM}. While the inelastic collisions only play a minor role for the elastic transport process studied here, our experimental means to distinguish them via mass spectrometry delivers intriguing insights into the stability of the ionic impurity in the condensate, as we discuss in more detail elsewhere \cite{Dieterle2020PRA}.

In conclusion, we have investigated transport of a single, untrapped low-energy ionic impurity through a BEC. The charged impurity is produced from a single Rydberg atom via a double-pulse field-ionization scheme minimizing its initial kinetic energy. We have observed indications of diffusive transport by the ion through the condensate via small electric bias fields and have measured the ion's mobility. The comparatively high BEC density in combination with the large ion-atom Langevin scattering cross section causes frequent collisions on a tens of microseconds timescale before the ion exits the gas.
While the transport dynamics studied in this work is well explained by semi-classical Langevin collisions, the energy scales accessed here are only limited by the level of stray field control. Improved control over the electric fields combined with high spatial resolution may allow in future experiments to enter a regime where only few partial waves contribute and quantum effects start to dictate the transport. It is very appealing to augment our techniques by ion-imaging optics, which is currently explored for ultracold Rydberg experiments, and  allows for sub-micrometer spatial resolution and excellent time resolution \cite{Stecker2017}. This provides exciting prospects to study spatially resolved ultracold ion-atom scattering from the level of individual collisions \cite{Schmid2018} to situations where collective phenomena play a role, comprising quantum effects for ion impurity transport and associated polaron-type dynamics \cite{Cote2000,Casteels2011,Astrakharchik2020}. 

We acknowledge support from Deutsche Forschungsgemeinschaft [Projects No. PF 381/13-1 and No. PF 381/17-1, the latter being part of the SPP 1929 (GiRyd)] and the Carl Zeiss Foundation via IQST. F. M. is indebted to the Baden-W\"urttemberg-Stiftung for the financial support by the Eliteprogramm for Postdocs. K.J. acknowledges support from the Polish National Agency for Academic Exchange (NAWA) via the Polish Returns 2019 programme.

%%%%%%%%%%%%%%-------------------References--------------------%%%%%%%%%%%%%%%%%%

\newpage
\section{Supplementary Material: Transport of a single cold ion immersed in a Bose-Einstein condensate}

\subsection{Rydberg excitation}
We implant a single low-energy \Rbplus ion into the central region of the condensate by exciting one atom from the atomic ensemble to a highly excited $160S_{1/2}$ Rydberg level using a $500 \, \rm{ns}$ long two-photon excitation pulse incorporating the intermediate $6P_{3/2}$ state at a detuning of $+320$ MHz. To account for density-induced level shifts of the Ryd\-berg line in the BEC, the lasers are detuned by -40 MHz from the bare Rydberg state energy \cite{Balewski2013MAT}. One of the excitation lasers is focused into the condensate center through a high-NA aspheric lens to a beam waist of about $1.8 \, \mu \rm{m}$, providing excellent spatial control over the position of the created Rydberg atom. Note that strong Rydberg blockade inhibits more than one excitation at a time \cite{Balewski2013MAT}, and is at the heart of creating a single ionic impurity only.

\subsection{Ion detection}

To extract the ion from the BEC after a variable evolution time $t$, we interrupt the transport dynamics with a two-step electric field pulse guiding the ions towards a microchannel plate (MCP) for detection. The extraction sequence consists of a first field step to $E_{\rm{ex,1}} = 0.3 \, \rm{V/cm}$ applied for $5 \, \mu \rm{s}$, followed by a second step to $E_{\rm{ex,2}} = 54 \, \rm{V/cm}$ for another $15 \, \mu \rm{s}$. Since $E_{\rm{ex,1}}$ is too small to field-ionize the precursor Rydberg atom, this scheme allows us to distinguish the rare cases where ion creation was not successful by their time of flight to the detector and to exclude them from the data analysis. In some cases, we detect a formed molecular ion \Rbmolplus, which due to its larger mass arrives at much later time at the MCP.

\subsection{Pulsed field ionization characterization}
\label{sec:pulsedionization}

\begin{figure}[!ht]
\centering
	\includegraphics[width=\columnwidth]{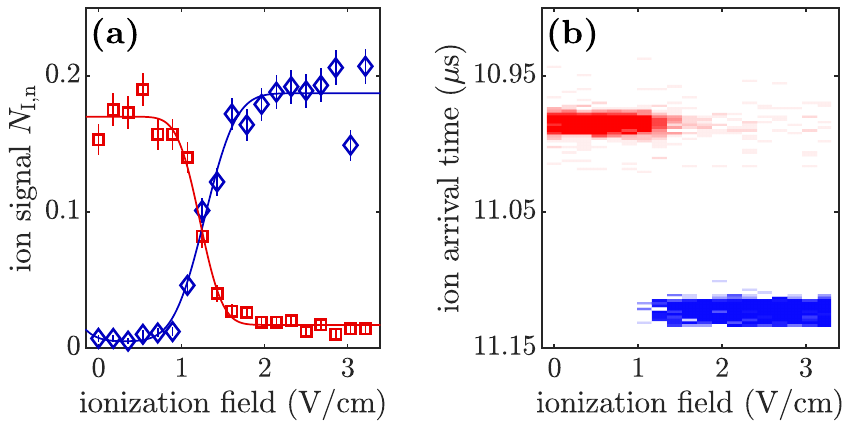}
	\caption{Pulsed field ionization (PFI) to produce a single ionic impurity. (a) \Rbplus ion (blue diamonds) and $160S_{1/2}$ Rydberg atom (red squares) signal resulting from the 10 ns long PFI as function of the height of the field ionization pulse. Solid lines are error-function fits to the data to guide the eye. (b) Temporal separation of ion and Rydberg atom signal on the MCP. Ions produced by the PFI arrive at 11.12 $\mu$s, whereas Rydberg atoms (ionized by $E_{\rm{ex,2}}$) arrive at 10.98 $\mu$s.}
	\label{Fig1SupMat}
\end{figure}

As mentioned in the main text, the low-energy ion is produced from a the single Rydberg atom employing a short (10 ns long) electric field-ionization pulse of $2.85 \, \rm{V/cm}$, followed by a second tailored field pulse of similar shape but with opposite polarity to counteract the rapid acceleration of the ion induced by the ionization field.

Here, we characterize the performance of the pulsed field ionization (PFI) sequence used to produce the low-energy ionic impurity. The following data are all obtained in a dilute thermal cloud with $n_{\rm{at}} \approx 1\times10^{13} \, \rm{cm}^{-3}$ to exclude any collision effects. First, the ionization efficiency of the precursor $160S_{1/2}$ Rydberg atom by the first 10 ns long ionization pulse is analyzed. In this measurement, we do not yet apply a second field pulse of opposite polarity. Fig.~\ref{Fig1SupMat}(b) shows the signal of produced \Rbplus (blue diamonds) and detected Rydberg atoms (red squares) as a function of the height of the field ionization pulse. We find that the $160S_{1/2}$ Rydberg level is ionized between $0.7$ and $2 \rm{V/cm}$ with an overall efficiency of $>90\%$ for $E \ge 2 \, \rm{V/cm}$. Instrumental for this analysis is the applied two-step electric field detection sequence, which allows us to temporally distinguish \Rbplus ions and Ryd\-berg atoms on the MCP [Fig.~\ref{Fig1SupMat}(b)]. Note that the latter are ionized by the field $E_{\rm{ex,2}}$ of the detection sequence.

\begin{figure}[!t]
\centering
	\includegraphics[width=\columnwidth]{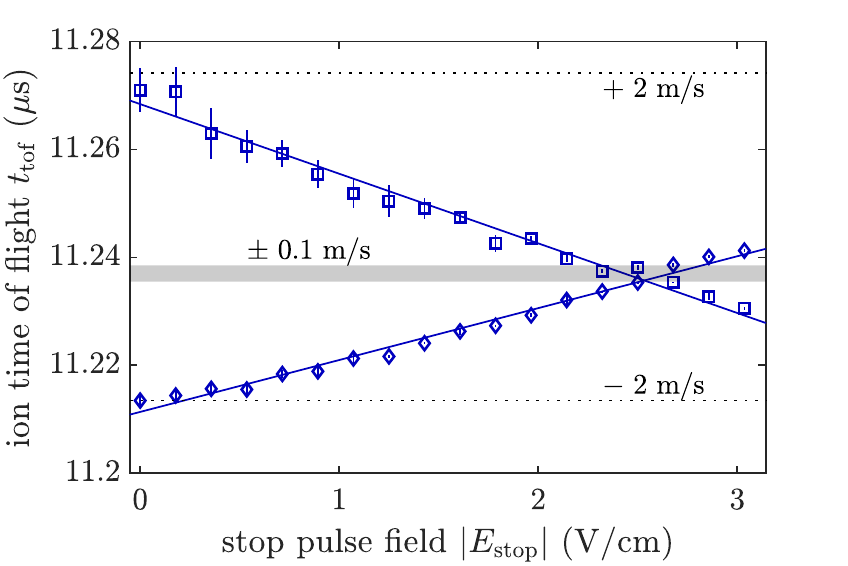}
	\caption{Characterization of the stopping procedure for minimizing the initial ion velocity. Mean time of flight $t_{\rm{tof}}$ of the ion to the MCP for $t =21 \, \mu$s and $E = \,4.3 \,\rm{mV/cm}$ as a function of the nominal height of the stop pulse. Squares (diamonds) show data for which the first pulse ($2.85 \, \rm{V/cm}$) of the PFI double-pulse sequence is aligned (anti-aligned) with the bias field $E$. Error bars for the experimental data originate from the standard deviation of the fitted center of the ion time of flight distribution to its center of mass. The shaded region and dotted lines indicate predicted values of $t_{\rm{tof}}$ for ions with initial velocity $\lesssim 0.1 \, \rm{m/s}$ and $\pm 2 \, \rm{m/s}$, respectively, as obtained from SIMION simulations. Solid lines are linear fits to the data and indicate the crossing point giving the optimal stop pulse field.} 
	\label{Fig2SupMat}
\end{figure}

Next, we add the second short electric field pulse of opposite polarity to the PFI sequence, which aims to minimize the \Rbplus velocity and thus produces a cold low-energy impurity. To quantify the performance of this stopping procedure, we let the ion evolve for $t = \,21 \mu$s in a bias electric field set to $E = 4.3 \,\rm{mV/cm}$ and monitor its time of flight $t_{\rm{tof}}$ to the MCP as a function of the nominal height of the second stop pulse (squares in Fig.~\ref{Fig2SupMat}). The change in $t_{\rm{tof}}$ already indicates the influence of the stop pulse. Specifically, smaller values of $t_{\rm{tof}}$ correspond to ion positions closer to the initial point of creation. However, this observation does not yet allow to identify the value of the second pulse height which leads to a minimal initial velocity. Therefore, we repeat the measurement but with the polarity of the two pulses reversed (diamonds in Fig.~\ref{Fig2SupMat}). This inverts the observed slope. The crossing point indicates a setting where the PFI sequence causes minimal velocity of the produced ion and thus allows for deducing the optimal value of the stop pulse height. For the experiments presented in the main text, the stop pulse height is thus chosen to $-2.66 \, \rm{V/cm}$. Moreover, a comparison of the data with our SIMION simulation allows us to quantify the ion's initial velocity to $\lesssim 0.1 \, \rm{m/s}$ (shaded region in Fig.~\ref{Fig2SupMat}). Note that the initial velocity of the ion in the absence of the stop pulse can be estimated in the same way to $\pm 2 \,\rm{m/s}$ depending on the polarity of the first field pulse (dotted lines in Fig.~\ref{Fig2SupMat}).

\subsection{SIMION simulations}
\label{sec:simion}

The quantitative analysis of the diffusive transport of the ion in the BEC involves simulations of our ion detection sequence using the charged-particle trajectory solver SIMION \cite{SimionMAT}. These simulations require a careful calibration taking into account the specific electric field and ion-lens configuration in our setup. First, the high voltage ($-2.35 \,\rm{kV}$) applied to the front plate of the MCP as well as to the last cylinder of the einzel lens leads to stray electric fields in the BEC region. Like in the experiment, such stray fields need to be compensated. In the simulation, we minimize these fields at the initial ion position by applying differential voltages to opposing field plates. Second, we use the low-density experimental data, i.e. the ballistic ion motion, for a fine calibration of the SIMION simulations. We find that we need to introduce two calibration parameters in order to reproduce the ion's time of flight distribution as a function of the evolution time $t$. Those are a gradient in the electric field and a small shift of the initial ion position with respect to the electrode center both along the $y$-axis (long axis of the BEC). The calibration procedure yields optimal agreement with the ballistic experimental data for a shift of $50 \, \mu\rm{m}$ and a field gradient of 1.18 mV/cm across 10$\mu$m. The field gradient is most likely a result of surface charges on the vacuum glass cell of our setup. Note that we have independently verified the presence and also the magnitude of the field gradient via Stark spectroscopy measurements for several positions of the Rydberg excitation along the $y$-axis.

\subsection{Stochastic trajectory simulations}
\label{sec:stochasticiondynamics}

The numerical simulations of the ion transport reported in the main text are based on Monte-Carlo sampled classical trajectories sequentially interrupted by ion-atom Langevin collisions, a model which was first developed to model buffer-gas cooling dynamics of a trapped ion \cite{Zipkes2011MAT}. Specifically, a single ion trajectory of the Monte-Carlo sample is computed assuming that the ion starts from its initial position ${\bf{x}}_{\rm{i}}^0$ and follows a classical trajectory in the applied electric field $E$. After a randomly sampled time $t_n$, a collision happens which changes the ion's current velocity ${\bf{v}}_{\rm{i}}^n$ instantaneously to a new value ${\bf{v}}_{\rm{i}}^{n+1}$. The elastic nature of the collision implies that, in the relative coordinate frame of a colliding ion-atom pair, the updated relative velocity ${\bf{v}}_{\rm{rel}}^{n+1} = R(\theta^n,\phi^n) {\bf{v}}_{\rm{rel}}^{n}$ \cite{Chen2014MAT}. Here, $R$ is a rotation matrix with polar angles $\theta^n$ and $\phi^n$. Accordingly, in the laboratory frame the ion velocity after the collision reads ${\bf{v}}_{\rm{i}}^{n+1} = 1/2 [{\bf{v}}_{\rm{i}}^{n} + R(\theta,\phi) {\bf{v}}_{\rm{i}}^{n}]$, where we have assumed that the colliding neutral atom is initially at rest, which is justified by the small BEC temperature compared to the kinetic energy of the ion. The relative collision angles $\theta^n$ and $\phi^n$ entering the rotation matrix are to be sampled according to the ion-atom differential cross section. For the spiraling-type Langevin collisions relevant for this work, $\theta^n$ and $\phi^n$ are drawn uniformly on the unit sphere, i.e. resembling the hard-sphere Langevin differential scattering cross section. Finally, the time steps $t_n$ between successive collisions are randomly sampled according to the energy-independent but density-dependent Langevin scattering rate $\gamma_{\rm{L}}=2 \pi n_{\rm{at}} \sqrt{C_4/\mu}$ \cite{Cote2000bMAT}. This allows us to account for the Thomas-Fermi density profile $n_{\rm{at}}(x,y,z)$ of the BEC with parameters given in the main text. Note that we also take care of the fact that $n_{\rm{at}}$ changes during the motion of the ion between collisions by applying the sampling algorithm for $t^n$ described in Ref.~\cite{Zipkes2011MAT}.

Finally, the results in the main text are obtained by typically 500 such trajectories. For the numerical data shown in Fig. 1(b) of the main text, the ion's initial position for each trajectory is the BEC center. For the numerical results which are compared to the experimental data in Fig. 3, Fig. 4(b) and (c), the ion's initial position is randomly sampled according to the Gaussian profile of our Rydberg excitation laser and the BEC density profile. These data sets also include averaging over small stray electric fields of $\pm 1 \, \rm{mV/cm}$ in all three directions. Incorporating these stray fields has practically no influence on the mobility simulations, but slightly broaden the distributions shown in Fig. 3.

Note that this treatment ignores the glancing-type collisions with small scattering angles. Indeed, these collisions barely affect the motion of the ion and do not contribute to diffusion or thermalization processes \cite{Cote2000bMAT}, but it is a priori not clear that they do not affect the transport dynamics in our finite size sample. For this reason, we have performed additional simulations where the scattering angles are sampled according to the full quantum mechanical ion-atom elastic scattering cross section derived from a random-phase approximation \cite{Zipkes2011MAT}. This elastic cross section features a pronounced forward scattering peak, indicative of the glancing collisions. Here, the time steps $t_n$ are sampled according to the total elastic cross section $\sigma_{\rm{el}}=\pi (\mu C_4^2)^{1/3} (1+\pi^2/16) E_{\rm{c}}^{-1/3}$, which accounts for the additional frequent glancing collisions ($E_{\rm{c}}$ is the collision energy) \cite{Cote2000bMAT}. Comparing the two approaches, we find no discernible difference and thus conclude that glancing collisions are not relevant for the dynamics we observe in our experiments.

\subsection{Molecular ion formation}
\label{sec:chemistry}

\begin{figure}[t]
\centering
	\includegraphics[width=\columnwidth]{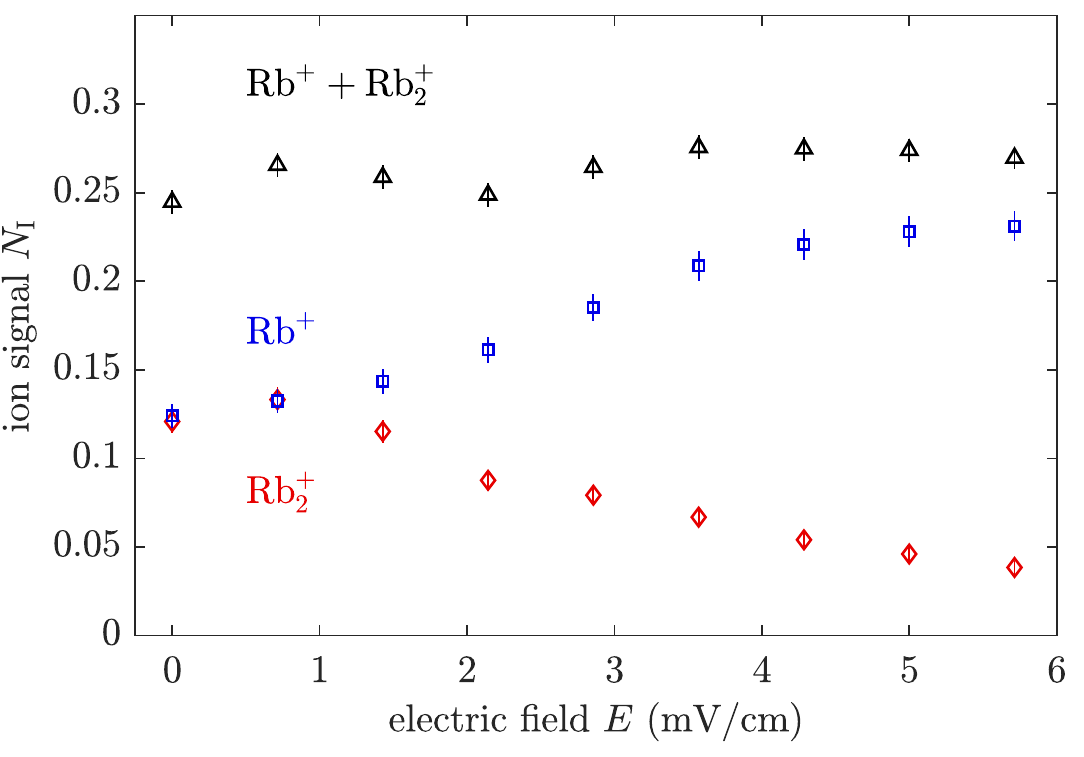}
	\caption{\Rbmolplus formation via three-body recombination. \Rbplus and \Rbmolplus signal as a function of the transport electric field $E$ after an evolution time $t = 21 \,\mu$s. Black triangles show the sum of the \Rbplus and \Rbmolplus signal.}
	\label{Fig3SupMat}
\end{figure}

Here, we briefly extend the discussion of the role of inelastic collisions, i.e. the formation of \Rbmolplus molecular ions via ion-atom-atom three-body recombination, for the measurement of the ion mobility, and refer to Ref.~\cite{Dieterle2020PRAMAT} for an in-depth study of the associated cold chemistry. Figure~\ref{Fig3SupMat} reveals the measured fraction of ions undergoing molecular ion formation as a function of $E$ for an evolution time of $t = 21 \,\mu\rm{s}$. We observe that for vanishing $E$ almost 50\% of the generated ions undergo three-body recombination. As explained in the main text, those events are excluded from the data analysis, leading to an effective reduction of the atomic density that we probe with the \Rbplus ions and thus qualitatively to a larger ion mobility. Indeed, by excluding the central $30 \, \%$ of the ions from our numerical simulations, we estimate that such an effect may alter $\mu_{\rm{ion}}^{\rm{sim}}$ by about $10 \, \%$.

%%%%%%%%%%%%%%-------------------References--------------------%%%%%%%%%%%%%%%%%%

\end{document}